\documentclass[%
twocolumn,
superscriptaddress,
showpacs, %preprintnumbers,
 amsmath,
 amssymb,
prc,
floatfix,
]{revtex4}

\usepackage{graphicx}% Include figure files
\usepackage{dcolumn}% Align table columns on decimal point
\usepackage{bm}% bold math
\usepackage{natbib}
\usepackage{color}
\begin{document}

\newcommand{\etal}{{\it et al.}}
\makeatother

\newcommand{\bra}[1]{\left\langle #1 \right|}
\newcommand{\brared}[1]{\langle #1 ||}
\newcommand{\ee}{\eta^{\ast},\eta}
\newcommand{\product}[2]{\left\langle #1 | #2 \right\rangle}

\newcommand{\kbar}{\bar{k}}
\newcommand{\ket}[1]{\left| #1 \right\rangle}
\newcommand{\ketred}[1]{|| #1 \rangle}
\newcommand{\ahat}{\hat{a}}
\newcommand{\adag}{a^{\dagger}}
\newcommand{\ahatdag}{\hat{a}^{\dagger}}
\newcommand{\Ahat}{\hat{A}}
\newcommand{\Adag}{A^{\dagger}}
\newcommand{\Ahatdag}{\hat{A}^{\dagger}}
\newcommand{\Atdag}{\hat{A}^{(\tau)\dagger}}
\newcommand{\Bdag}{\hat{B}^{\dagger}}
\newcommand{\Bhat}{\hat{B}}
\newcommand{\Bhatdag}{\hat{B}^{\dagger}}
\newcommand{\Btdag}{\hat{B}^{(\tau)\dagger}}
\newcommand{\bhat}{\hat{b}}
\newcommand{\bdag}{b^{\dagger}}
\newcommand{\cdag}{c^{\dagger}}
\newcommand{\chat}{\hat{c}}
\newcommand{\chatdag}{\hat{c}^{\dagger}}
\newcommand{\degree}{^{\circ}}
\newcommand{\sprime}{s^{\prime}}
\newcommand{\Hhat}{\hat{H}}
\newcommand{\Hhatp}{\hat{H}^{\prime}}
\newcommand{\Ihat}{\hat{I}}
\newcommand{\Jhat}{\hat{J}}
\newcommand{\hhat}{\hat{h}}
\newcommand{\fp}{f^{(+)}}
\newcommand{\fpp}{f^{(+)\prime}}
\newcommand{\fm}{f^{(-)}}
\newcommand{\Fhat}{\hat{F}}
\newcommand{\Fhatdag}{\hat{F}^\dagger}
\newcommand{\Fhatp}{\hat{F}^{(+)}}
\newcommand{\Fhatm}{\hat{F}^{(-)}}
\newcommand{\Fhatpm}{\hat{F}^{(\pm)}}
\newcommand{\Fhatdagpm}{\hat{F}^{\dagger(\pm)}}
\newcommand{\Hc}{{\cal H}}
\newcommand{\Hcp}{{\cal H}^{\prime}}
\newcommand{\Ic}{{\cal I}}
\newcommand{\It}{\widetilde{I}}
\newcommand{\ITV}{{\cal I}_{\rm TV}}
\newcommand{\Jc}{{\cal J}}
\newcommand{\jp}{j^{\prime}}
\newcommand{\Qc}{{\cal Q}}
\newcommand{\Pc}{{\cal P}}
\newcommand{\Ec}{{\cal E}}
\newcommand{\Sc}{{\cal S}}
\newcommand{\Rc}{{\cal R}}

\newcommand{\ddg}{d^{\dagger}}

\newcommand{\Nhat}{\hat{N}}
\newcommand{\Nt}{\widetilde{N}}
\newcommand{\Vt}{\widetilde{V}}
\newcommand{\nL}[1]{n_{L_{#1}}}
\newcommand{\nK}[1]{n_{K_{#1}}}
\newcommand{\nKb}{\mbox{\boldmath $n_K$}}
\newcommand{\nLb}{\mbox{\boldmath $n_L$}}

\newcommand{\mubar}{\bar{\mu}}

\newcommand{\Dc}{{\mathscr D}}
\newcommand{\Dp}{D^{(+)}}
\newcommand{\Ddag}{\hat{D}^{\dagger}}
\newcommand{\dhat}{\hat{d}}
\newcommand{\Dhat}{\hat{D}}
\newcommand{\Dhatp}{\hat{D}^{(+)}}
\newcommand{\Ghat}{\hat{G}}
\newcommand{\Glambda}{G^{(\lambda)}}
\newcommand{\Gstarlambda}{G^{(\lambda)\ast}}
\newcommand{\Qhat}{\hat{Q}}
\newcommand{\Rhat}{\hat{R}}
\newcommand{\Phat}{\hat{P}}
\newcommand{\Pdag}{\hat{P}^{\dagger}}
\newcommand{\Psihat}{\hat{\Psi}}
\newcommand{\Qdag}{Q^{\dagger}}
\newcommand{\That}{\hat{\Theta}}
\newcommand{\Thatt}{\widetilde{\hat{\Theta}}}
\newcommand{\Tr}{{\rm Tr}}

\newcommand{\ktilde}{\tilde{k}}

\newcommand{\Pcirc}{\stackrel{\circ}{P}}
\newcommand{\Qcirc}{\stackrel{\circ}{Q}}
\newcommand{\Ncirc}{\stackrel{\circ}{N}}
\newcommand{\Tcirc}{\stackrel{\circ}{\Theta}}
\newcommand{\Pcircp}{\stackrel{\circ}{P^{\prime}}}
\newcommand{\Qcircp}{\stackrel{\circ}{Q^{\prime}}}
\newcommand{\Ncircp}{\stackrel{\circ}{N^{\prime}}}
\newcommand{\Tcircp}{\stackrel{\circ}{\Theta^{\prime}}}
\newcommand{\Fp}{F^{(+)}}
\newcommand{\Fm}{F^{(-)}}
\newcommand{\Rp}{R^{(+)}}
\newcommand{\Rm}{R^{(-)}}
\newcommand{\Bt}{\widetilde{B}}
\newcommand{\lambdat}{\widetilde{\lambda}}
\newcommand{\Phatt}{\widetilde{\hat{P}}}
\newcommand{\ab}{\bf a}

\newcommand{\Ab}{\mbox{\boldmath $A$}}
\newcommand{\Abdag}{\mbox{\boldmath $A$}^{\dagger}}
\newcommand{\Bb}{\mbox{\boldmath $B$}}
\newcommand{\cb}{\bf c}
\newcommand{\Db}{\mbox{\boldmath $D$}}
\newcommand{\Nb}{\mbox{\boldmath $N$}}
\newcommand{\Nbhat}{\hat{\mbox{\boldmath $N$}}}
\newcommand{\Qb}{\mbox{\boldmath $Q$}}
\newcommand{\Qhatt}{\widetilde{\hat{Q}}}
\newcommand{\Pb}{\mbox{\boldmath $P$}}
\newcommand{\phit}{\phi(t)}
\newcommand{\pdot}{\dot{p}}
\newcommand{\phix}[1]{\phi(#1)}
\newcommand{\qdot}{\dot{q}}
\newcommand{\phivib}{\phi(\eta^{\ast},\eta)}
\newcommand{\Ts}{{\cal T}}
\newcommand{\del}{\partial}
\newcommand{\eps}{\epsilon}
\newcommand{\beq}{\begin{equation}}
\newcommand{\beqa}{\begin{eqnarray}}
\newcommand{\eeq}{\end{equation}}
\newcommand{\eeqa}{\end{eqnarray}}
\newcommand{\Yb}{${}^{168}$Yb\ }
\newcommand{\Zhat}{\hat{Z}}
\newcommand{\rhodot}{\dot{\rho}}
\newcommand{\Khat}{\hat{K}}
\newcommand{\Kp}{K^{+}}
\newcommand{\Km}{K^{-}}
\newcommand{\Kz}{K^0}

\newcommand{\lb}{\bf l}
\newcommand{\sbold}{\bf s}

\newcommand{\Lp}{L^{+}}
\newcommand{\Lm}{L^{-}}
\newcommand{\Lz}{L^0}

\newcommand{\Mc}{{\cal M}}
\newcommand{\Mchat}{\hat{\cal M}}

\newcommand{\ddeta}{\frac{\partial}{\partial \eta}}
\newcommand{\ddetastar}{\frac{\partial}{\partial \eta^\ast}}
\newcommand{\etastar}{\eta^\ast}
\newcommand{\ketvib}{\ket{\phi (\etastar, \eta)}}
\newcommand{\bravib}{\bra{\phi (\etastar, \eta)}}
\newcommand{\zhateta}{\hat{z}(\eta)}
\newcommand{\zhat}{\hat{z}}
\newcommand{\oo}{\stackrel{\circ}{O}(\etastar,\eta)}
\newcommand{\oodag}{\stackrel{\circ}{O^{\dagger}}(\etastar,\eta)}
\newcommand{\oodagp}{\stackrel{\circ}{O^{\dagger\prime}}(\etastar,\eta)}
\newcommand{\oop}{\stackrel{\circ}{O^{\prime}}(\etastar,\eta)}
\newcommand{\Odag}{\hat{O}^{\dagger}}
\newcommand{\Ohat}{\hat{O}}
\newcommand{\Uinv}{U^{-1}(\etastar, \eta)}
\newcommand{\Uinvp}{U^{-1}(\etastar,\eta,\varphi,n)}
\newcommand{\U}{U(\etastar, \eta)}
\newcommand{\Up}{U(\etastar,\eta,\varphi,n)}
\newcommand{\etader}{\frac{\del}{\del \eta}}
\newcommand{\etastarder}{\frac{\del}{\del \etastar}}

\newcommand{\fb}{\mbox {\bfseries\itshape f}}
\newcommand{\SB}{\mbox {\bfseries\itshape S}}

\newcommand{\vbar}{\bar{v}}

\newcommand{\Udag}{U^{\dagger}}
\newcommand{\Vdag}{V^{\dagger}}

\newcommand{\Wc}{{\cal W}}
\newcommand{\Wcdag}{{\cal W}^{\dagger}}

\newcommand{\Xhat}{\hat{X}}
\newcommand{\Xdag}{\hat{X}^{\dagger}}

\renewcommand{\thanks}{\footnote}
\newcommand\tocite[1]{$^{\hbox{--}}$\cite{#1}}%\cite{xx}\tocite{yy}

\preprint{}

\title{
Shape fluctuations in the ground and excited $0^+$ states \\ of $^{30}$Mg  and $^{32}$Mg
}% Force line breaks with \\
%\thanks{A footnote to the article title}%

\author{Nobuo Hinohara}
%\email{hinohara@riken.jp}
\author{Koichi Sato}%
\affiliation{%
Theoretical Nuclear Physics Laboratory, RIKEN Nishina Center, Wako 351-0198, Japan
}%
\author{Kenichi Yoshida}
\affiliation{%
Department of Physics, Faculty of Science, Niigata University, Niigata 950-2181, Japan
}%
\affiliation{%
Theoretical Nuclear Physics Laboratory, RIKEN Nishina Center, Wako 351-0198, Japan
}%
\author{Takashi Nakatsukasa}
\affiliation{%
Theoretical Nuclear Physics Laboratory, RIKEN Nishina Center, Wako 351-0198, Japan
}%
\author{Masayuki Matsuo}
\affiliation{%
Department of Physics, Faculty of Science, Niigata University, Niigata 950-2181, Japan
}%
\author{Kenichi Matsuyanagi}
\affiliation{%
Theoretical Nuclear Physics Laboratory, RIKEN Nishina Center, Wako 351-0198, Japan
}%
\affiliation{%
Yukawa Institute for Theoretical Physics, Kyoto University, Kyoto 606-8502, Japan
}%

\date{\today}% It is always \today, today,
             %  but any date may be explicitly specified

\begin{abstract}
Large-amplitude collective dynamics of shape phase transition
in the low-lying states of $^{30-36}$Mg is investigated by solving 
the five-dimensional (5D) quadrupole collective Schr\"odinger equation. 
The collective masses and potentials of the 5D collective Hamiltonian are
microscopically derived with use of the constrained Hartree-Fock-Bogoliubov 
plus local quasiparticle RPA method.
Good agreement with the recent experimental data is obtained
for the excited $0^+$ states as well as the ground bands.
For $^{30}$Mg, the shape coexistence picture 
that the deformed excited $0^+$ state coexists with the spherical ground state 
approximately holds. 
On the other hand, large-amplitude quadrupole-shape fluctuations 
dominate in both the ground and the excited $0^+$ states in $^{32}$Mg, 
so that the interpretation of `coexisting spherical excited $0^+$ state'   
based on the naive inversion picture of the spherical and deformed configurations 
does not hold.
\end{abstract}

\pacs{21.60.Ev, 21.10.Re, 21.60.Jz, 27.30.+t}% PACS, the Physics and Astronomy
                             % Classification Scheme.
%\keywords{Suggested keywords}%Use showkeys class option if keyword
                              %display desired
\maketitle

%\tableofcontents

Nuclei exhibit a variety of shapes in their ground and excited states. A remarkable
feature of the quantum phase transition of a finite system is that the order
parameters (shape deformation parameters) always fluctuate and vary with the particle
number. Especially, the large-amplitude shape fluctuations play a crucial role in
transitional (critical) regions. Spectroscopic studies of low-lying excited states
in transitional nuclei are of great interest to observe such unique features of the
finite quantum systems.

Low-lying states of neutron-rich nuclei
around $N=20$ attract a great interest, 
as the spherical configurations associated with the magic number
disappear in the ground states. 
In neutron-rich Mg isotopes,
the increase of the excitation energy ratio $E(4_1^+)/E(2_1^+)$  
\cite{PhysRevC.82.034305,PhysRevC.79.054319, Yoneda2001233}
and the enhancement of $B(E2;2^+_1\to 0^+_1)$ from $^{30}$Mg to $^{34}$Mg  
\cite{PhysRevLett.94.172501,Motobayashi19959,Iwasaki2001227}
indicate a kind of quantum phase transition 
from spherical to deformed shapes taking place around $^{32}$Mg. 
These experiments stimulate microscopic investigations on
quadrupole collective dynamics unique to this region
of the nuclear chart with various theoretical approaches; the shell model 
\cite{PhysRevC.41.1147, PhysRevC.60.054315, 
Caurier2001374,springerlink:10.1140/epja/i2003-10201-5}, 
the Hartree-Fock-Bogoliubov (HFB) method 
\cite{Terasaki1997706, PhysRevC.60.014316},
the parity-projected HF
\cite{EPJA25s01_549}, 
the quasiparticle RPA (QRPA) \cite{PhysRevC.69.034301,yoshida:044312}, 
the angular-momentum projected generator coordinate method (GCM) 
with \cite{RodriguezGuzman2002201} and 
without \cite{PhysRevC.83.014308, IJMPE20_482}
restriction to the axial symmetry,  
and the antisymmetrized molecular dynamics \cite{PTP.107.33}. 

Quite recently, excited $0^+$ states were found in $^{30}$Mg 
\cite{springerlink:10.1140/epjad/i2005-06-159-0,PhysRevLett.103.012501}
and $^{32}$Mg \cite{PhysRevLett.105.252501} at 1.789 MeV and 1.058 MeV, respectively.
For $^{30}$Mg, the excited $0_2^+$ state is interpreted as a prolately deformed
state which coexists with the spherical ground state. 
For $^{32}$Mg, from the observed population of the excited $0_2^+$ state in the $(t,p)$ reaction on $^{30}$Mg,
it is suggested \cite{PhysRevLett.105.252501} that the
$0_2^+$ state is a spherical state coexisting with the deformed ground state and that 
their relative energies are inverted at $N=20$.
However, available shell-model and GCM calculations considerably overestimate 
its excitation energy ($1.4-3.1$ MeV) 
\cite{Caurier2001374,springerlink:10.1140/epja/i2003-10201-5,
RodriguezGuzman2002201,PhysRevLett.103.012501}.
It is therefore a challenge for modern microscopic theories of nuclear structure 
to clarify the nature of the excited $0_2^+$ states. 
For understanding shape dynamics in low-lying collective excited states 
of Mg isotopes near $N=20$, it is certainly desirable to develop a theory 
capable of describing various situations in a unified manner, 
including, at least, 1) an ideal shape coexistence limit where 
the wave function of an individual quantum state 
is well localized in the deformation space 
and 2) a transitional situation 
where the large-amplitude shape fluctuations dominate.   

In this article, we microscopically derive the five-dimensional (5D) 
quadrupole collective Hamiltonian 
using the constrained Hartree-Fock-Bogoliubov (CHFB)
plus local QRPA (LQRPA) method 
\cite{PhysRevC.82.064313}.
The 5D collective Hamiltonian takes into account 
all the five quadrupole degrees of freedom:
the axial and triaxial quadrupole deformations $(\beta, \gamma)$ 
and the three Euler angles.
This approach is suitable for our purpose of describing a variety of 
quadrupole collective phenomena in a unified way.  
Another advantage is that the time-odd mean-field contributions are taken into account
in evaluating the vibrational and rotational inertial functions. 
In spite of their importance for correctly describing collective excited states, 
the time-odd contributions are ignored in the widely used 
Inglis-Belyaev cranking formula for inertial functions.  
The CHFB + LQRPA method has been successfully applied to various large-amplitude 
collective dynamics including the oblate-prolate shape coexistence phenomena 
in Se and Kr isotopes \cite{PhysRevC.82.064313,Sato201153}, 
the $\gamma$-soft dynamics in $sd$-shell nuclei \cite{PhysRevC.83.014321}, 
and the shape phase transition in neutron-rich Cr isotopes \cite{PhysRevC.83.061302}.
A preliminary version of this work was reported in Ref.~\cite{Hinohara:2011fv}.

The 5D quadrupole collective Hamiltonian is written as
\begin{align}
 {\cal H}_{\rm coll} &= T_{\rm vib} + T_{\rm rot} + V(\beta,\gamma), \label{eq:collH}\\
 T_{\rm vib} &= \frac{1}{2}D_{\beta\beta}(\beta,\gamma) \dot{\beta}^2
+ D_{\beta\gamma}(\beta,\gamma)\dot{\beta}\dot{\gamma}
+ \frac{1}{2} D_{\gamma\gamma}(\beta,\gamma) \dot{\gamma}^2, \\
 T_{\rm rot} &= \frac{1}{2} \sum_{k=1}^3 {\cal J}_k(\beta,\gamma) \omega_k^2, 
\end{align}
where $T_{\rm vib}$ and $T_{\rm rot}$ are the vibrational and rotational 
kinetic energies, respectively, and $V$ is the collective potential.
The vibrational collective masses, $D_{\beta\beta}, D_{\beta\gamma}$, and  $D_{\gamma\gamma}$,
are the inertial functions for the $(\beta,\gamma)$ coordinates.
The rotational moments of inertia ${\cal J}_k$ associated with the 
three components of the rotational angular velocities $\omega_k$
are defined with respect to the principal axes.
In the CHFB + LQRPA method, 
the collective potential is calculated 
with the CHFB equation with four constraints on  
the two quadrupole operators and the proton and neutron numbers.  
The inertial functions in the collective Hamiltonian are determined
from the LQRPA normal modes
locally defined for each CHFB state in the $(\beta,\gamma)$ plane.
The equations 
to find the local normal modes are similar to the well-known QRPA equations,
but the equations are solved on top of the non-equilibrium CHFB states.  
Two LQRPA solutions 
representing quadrupole shape motion
are selected for the calculation of the vibrational 
inertial functions.
After quantizing the collective Hamiltonian (\ref{eq:collH}), 
we solve the 5D collective Schr\"odinger equation 
and obtain collective wave functions 
\begin{align}
 \Psi_{\alpha IM}(\beta,\gamma,\Omega) = \sum_{K={\rm even}}
\Phi_{\alpha IK}(\beta,\gamma) \langle\Omega|IMK\rangle,
\end{align}
where $\Phi_{\alpha IK}(\beta,\gamma)$ are the vibrational wave functions 
and $\langle\Omega|IMK\rangle$ are the rotational wave functions
defined in terms of ${\cal D}$ functions ${\cal D}^I_{MK}(\Omega)$.
We then evaluate
$E2$ matrix elements.  
More details of this approach are given in Ref.~\cite{PhysRevC.82.064313}.

% figure 1: collective potential

\begin{figure*}[t]
\begin{center}
\begin{tabular}{ccccl}
\includegraphics[height=38mm]{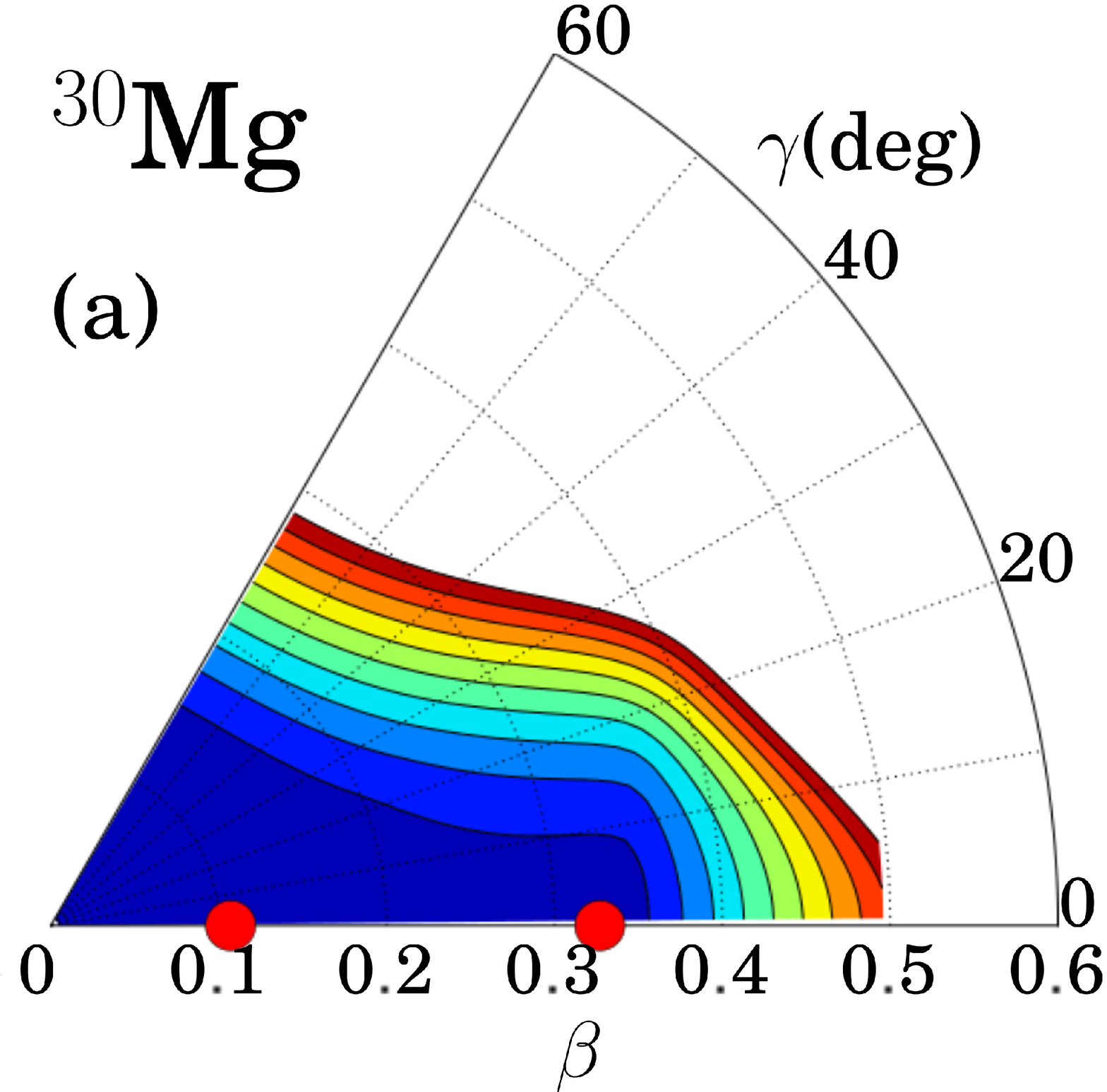} &
\includegraphics[height=38mm]{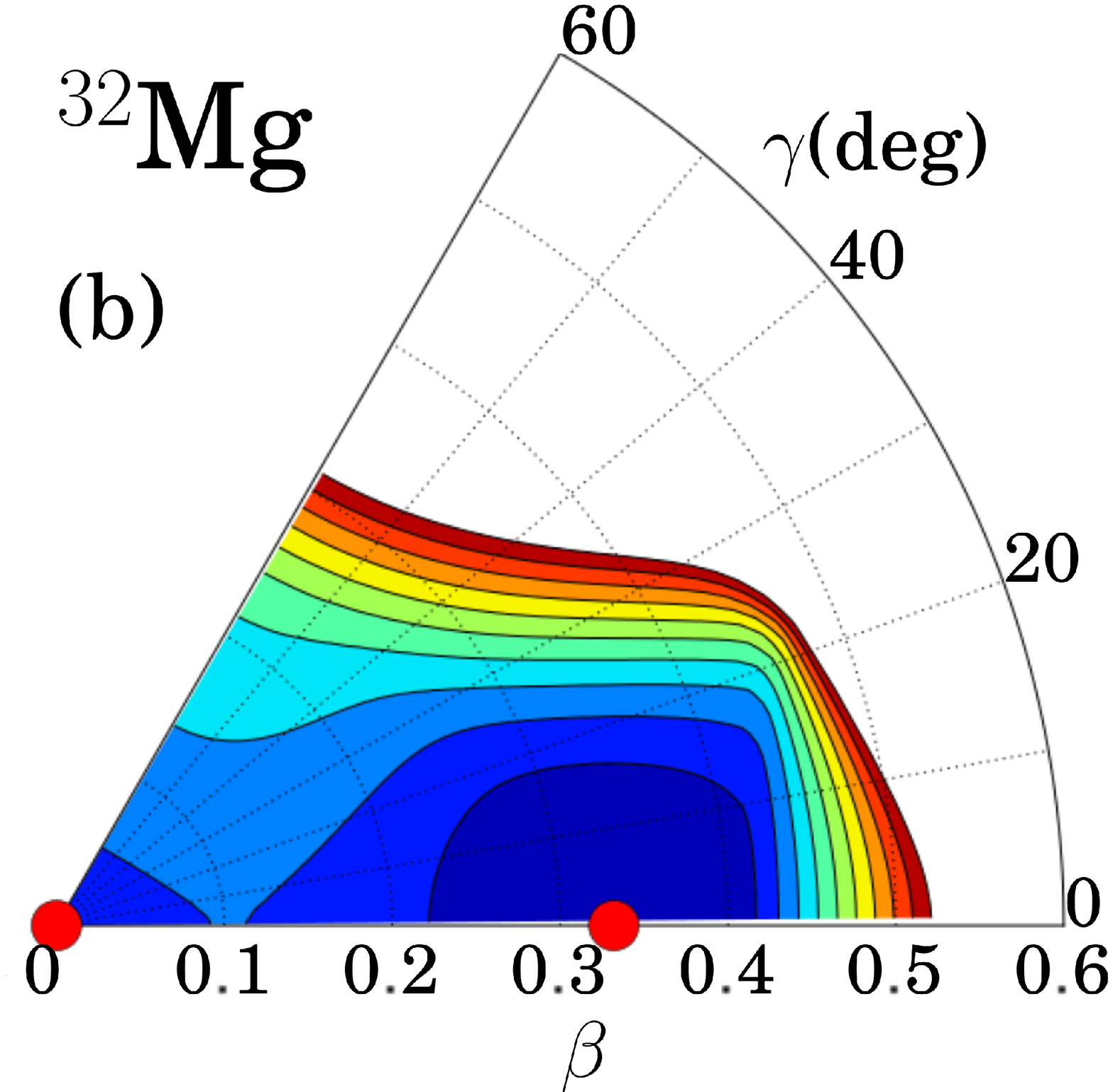} &
\includegraphics[height=38mm]{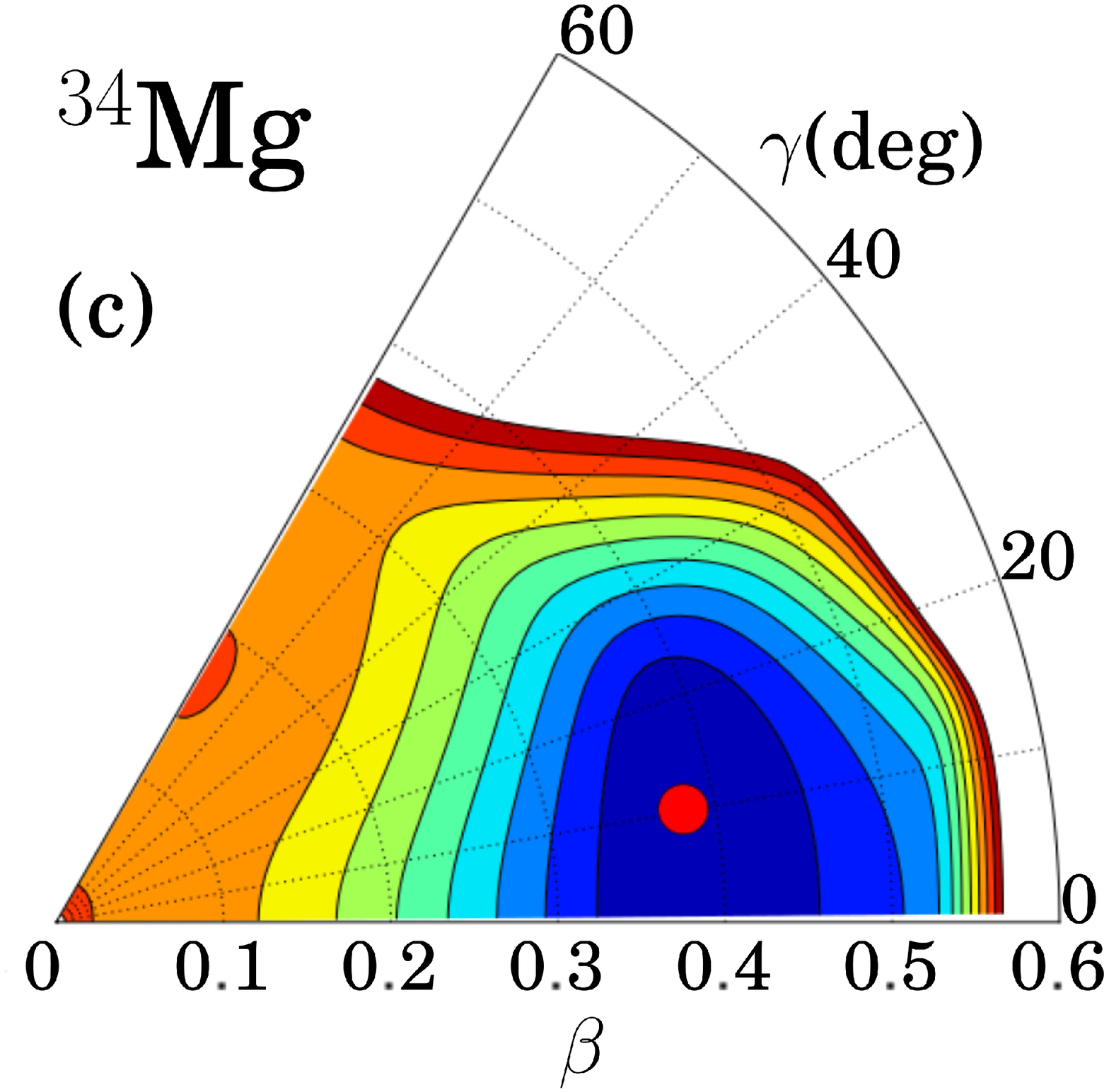} &
\includegraphics[height=38mm]{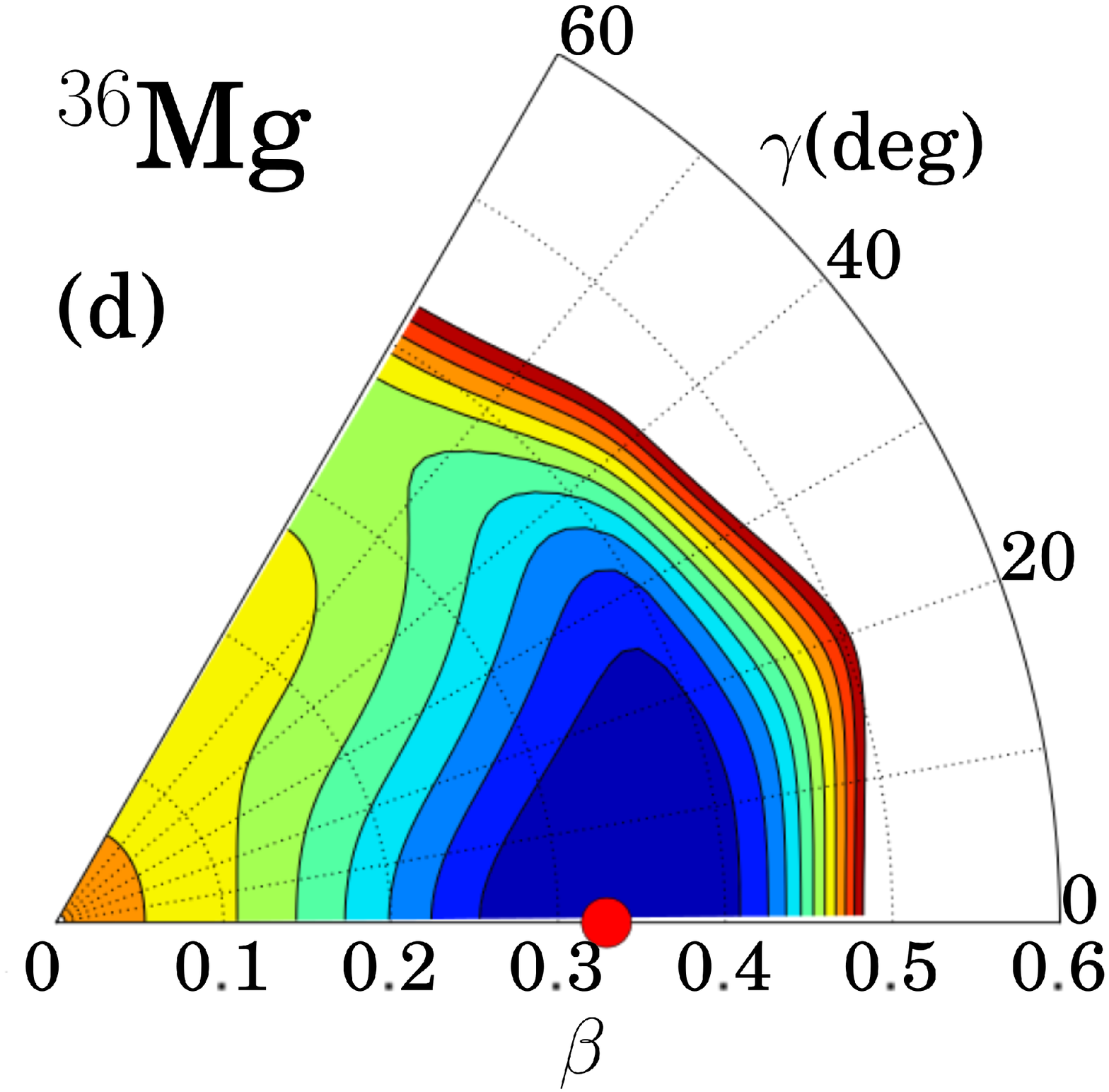} &
\includegraphics[height=38mm]{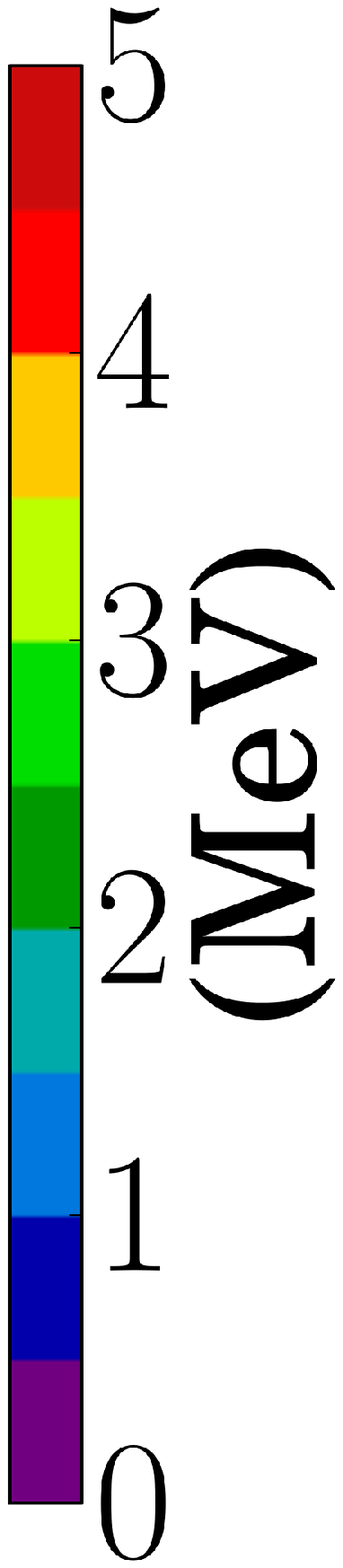}
\end{tabular}
\end{center}
\caption{ \label{fig:V} (Color online)
Collective potentials for $^{30-36}$Mg.
The HFB equilibrium points are indicated by red circles.}
\end{figure*}

We solve the CHFB + LQRPA equations employing, as a microscopic Hamiltonian, 
the pairing-plus-quadrupole (P+Q) model including the quadrupole-pairing interaction. 
As an active model space, the two-major harmonic oscillator shells ($sd$ and $pf$ shells)
are taken into account for both neutrons and protons.  
To determine the parameters in the P+Q Hamiltonian, 
we first perform Skyrme-HFB calculations 
with the SkM* functional and the surface pairing functional 
using the HFBTHO code \cite{Stoitsov200543}.
The pairing strength ($V_0=-374$ MeV fm$^{-3}$, with a cutoff quasiparticle
energy of 60 MeV) is fixed so as to reproduce the
experimental neutron gap of $^{30}$Ne (1.26 MeV). 
We then determine the parameters for each nucleus in the following way.
The single particle energies are determined by means of  
the constrained Skyrme-HFB calculation at the spherical shape. 
The resulting single particle energies (in the canonical basis) are then scaled 
with the effective mass of the SkM* functional $m^*/m=0.79$,  
since the P+Q model is designed to be used for single-particle 
states whose effective mass is equal to the bare nucleon mass. 
In $^{32}$Mg, the $N=20$ shell gap between $d_{3/2}$ and $f_{7/2}$  
is 3.7 MeV for the SkM* functional,
and it becomes 2.9 MeV after the effective mass scaling.
This value is appreciably smaller than the standard modified oscillator 
value 4.5 MeV \cite{Bengtsson198514}.  
This spacing almost stays constant for $^{30-36}$Mg.
The strengths of the monopole-pairing interaction are determined to 
reproduce the pairing gaps obtained in the Skyrme-HFB calculations 
at the spherical shape.
The strength of the quadrupole particle-hole interaction is determined  
to reproduce the magnitude of the axial quadrupole deformation $\beta$ 
of the Skyrme-HFB minimum.
The strengths of the quadrupole-pairing interaction are determined  
so as to fulfill the self-consistency condition \cite{Sakamoto1990321}.
We use the quadrupole polarization charge $\delta e_{\rm pol} = 0.5$
for both neutrons and protons when evaluating $E2$ matrix elements.
We solve the CHFB + LQRPA equations at 3600 $\beta$-$\gamma$ mesh points 
in the region $0<\beta<\beta_{\rm max}$ and $0^\circ<\gamma<60^\circ$,
with $\beta_{\rm max}=0.5$ for $^{30}$Mg and 0.6 for $^{32,34,36}$Mg.

Our theoretical framework is quite general and it can be used in conjunction with
various Skyrme forces/modern density functionals going beyond the P+Q model.
Then the effects of weakly bound neutrons and coupling to the continuum
on the properties of the low-lying collective excitations,
discussed in Refs.~\cite{PhysRevC.69.034301,yoshida:044312}, can 
be taken into account, for example, by solving the CHFB + LQRPA 
equations in the 3D coordinate mesh representation.
However, it requires a large-scale calculation with modern parallel 
processors and it remains as a challenging future subject.
A step toward this goal has recently been carried out for axially symmetric cases
\cite{PhysRevC.83.061302}.

Figure~\ref{fig:V} shows the collective potentials $V(\beta,\gamma)$ for 
$^{30-36}$Mg. It is clearly seen that 
prolate deformation grows with increase of the neutron number.
The collective potential for $^{30}$Mg is very soft with respect to $\beta$.
It has a minimum at $\beta=0.11$ and 
a local minimum at $\beta=0.33$.
The barrier height between the two minima is
only 0.24 MeV (measured from the lower minimum).
In $^{32}$Mg,  in addition to the prolate minimum 
at $\beta=0.33$, a spherical local minimum 
(associated with the $N=20$ spherical shell gap) appears. 
The barrier height between the two minima is 1.0 MeV 
(measured from the lower minimum).
The spherical local minimum disappears in $^{34}$Mg and $^{36}$Mg,
and the prolate minima become soft in the direction of triaxial deformation $\gamma$.
In $^{34}$Mg, the potential minimum is located at $\gamma=10^\circ$.

% figure 2: ground band properties
 
\begin{figure}[htbp]
\includegraphics[width=70mm]{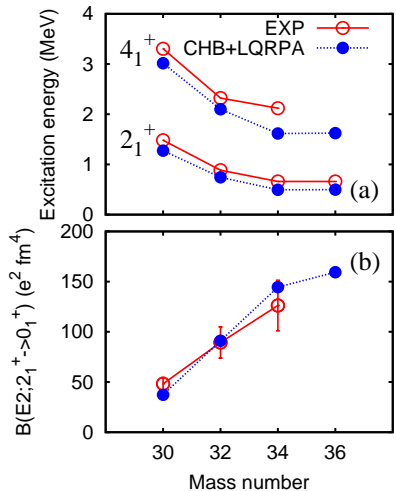}
\caption{ \label{fig:yrast} (Color online)
Comparison of calculated excitation energies of the $2_1^+$ and $4_1^+$ states 
(upper panel) and $B(E2;2_1^+\to 0_1)$ values (lower panel) in $^{30-36}$Mg 
with experimental data 
\cite{PhysRevC.82.034305,PhysRevC.79.054319, 
Yoneda2001233,PhysRevLett.94.172501,Motobayashi19959,Iwasaki2001227}.
}
\end{figure}
In Fig.~\ref{fig:yrast}, 
calculated excitation energies and $E2$ transition strengths are 
compared with the experimental data. 
The lowering of the excitation energies of the $2_1^+$ and  $4_1^+$ states
and the remarkable increase of $B(E2;2_1^+\to 0_1^+)$ 
from $^{30}$Mg to $^{34}$Mg are well described in this calculation.
The calculated ratio of the excitation energies, $E(4_1^+)/E(2_1^+)$, increases as 
2.37, 2.82, 3.26, and 3.26,  while the ratio of the transition strengths, 
$B(E2;4_1^+ \to 2_1^+)/B(E2;2_1^+ \to 0_1^+)$, decreases as 2.03, 1.76, 1.43, and 1.47, 
in going from $^{30}$Mg to $^{36}$Mg. 
Thus, the properties of the $2_1^+$ and $4_1^+$states 
gradually change from vibrational to rotational with increasing neutron number. 

% figure 3: excited band properties

\begin{figure}[htbp]
\includegraphics[width=70mm]{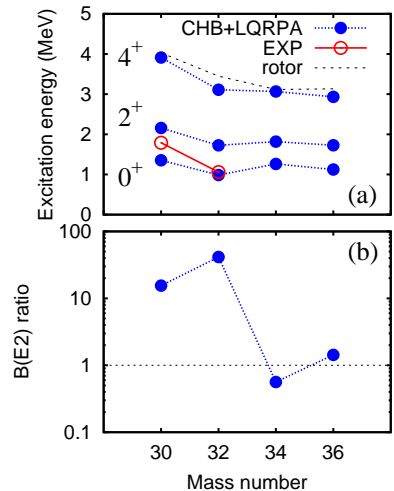}
\caption{ \label{fig:excited} (Color online)
Excitation energies of the excited $0_2^+$,  $2_{2,3}^+$ and $4_{2,3}^+$ states 
(upper panel) 
and the ratio $B(E2;0_2^+\to 2_1^+)/B(E2;0_1^+\to 2^+_{2,3})$ 
of the inter-band $E2$ transition strengths
between the $K=0_2$ and $K=0_1$ bands (lower panel).  
Experimental data are taken from 
Refs.~\cite{PhysRevLett.103.012501,PhysRevLett.105.252501}.
See texts for details.}
\end{figure}

Let us next discuss the properties of the $0^+_2$ states 
and the $2^+$ and $4^+$ states 
connected to the $0_2^+$ states with strong $E2$ transitions.
The result of calculation is presented in Fig.~\ref{fig:excited} 
together with the recent experimental data.  
The calculated excitation energies of the $0_2^+$ states
are 1.353 and 0.986 MeV for $^{30}$Mg and $^{32}$Mg, respectively, 
in fair agreement with the experimental data  
\cite{PhysRevLett.103.012501,PhysRevLett.105.252501}.
In particular, the very low excitation energy of the $0_2^+$ state in  
$^{32}$Mg is well reproduced.
In our calculation, more than 90\% (80\%) of the collective wave functions 
for the yrast  (excited) band members are 
composed of the $K=0$ component.  
Therefore we denote the ground band by `the $K=0_1$ band,' 
and the excited band by `the $K=0_2$ band.' 
The $2^+$ and $4^+$ states belonging to the $K=0_2$ band
appear as the second $2^+$ and $4^+$ states in $^{30,32}$Mg, 
while they appear as the third $2^+$ and $4^+$ states in $^{34,36}$Mg.  
Accordingly, we use $2^+_{2,3}$ and $4^+_{2,3}$, 
to collectively indicate the second or the third $2^+$ and $4^+$ states.
The calculated ratios of the excitation energies relative to the excited $0_2^+$ state, 
$[E(4_{2,3}^+)-E(0_2^+)]/[E(2_{2,3}^+)-E(0_2^+)]$, are 
3.18, 2.87, 3.25, and 3.00, 
for $^{30}$Mg, $^{32}$Mg, $^{34}$Mg, and $^{36}$Mg, respectively.  
In the upper panel of Fig.~\ref{fig:excited} 
we also plot the rotor-model prediction for the excitation energies of the $4^+$ states
estimated from the $0^+ -2^+$ spacings in the $K=0_2$ bands.
The deviation from the rotor-model prediction is largest in $^{32}$Mg 
indicating importance of shape-fluctuation effects. 
Although the calculated excitation spectrum of the $K=0_2$ band in $^{30}$Mg 
looks rotational, we find a significant deviation from the rotor-model prediction  
in the $E2$ transition properties. 
The calculated ratios of the $E2$ transition strengths, 
$B(E2;4_{2,3}^+\to 2_{2,3}^+)/B(E2;2_{2,3}^+\to 0^+_2)$, are 
1.05, 1.54, 1.47, and 1.51, 
for $^{30-36}$Mg, respectively. 
The deviation from the rotor-model value (1.43) is
largest in $^{30}$Mg.  
The significant deviation from the simple rotor-model pattern  
of the $K=0_2$ bands in $^{30}$Mg and $^{32}$Mg, noticed above,  
can be seen more drastically in the inter-band $E2$ transition properties. 
In the lower panel of Fig.~\ref{fig:excited}, we plot the ratio  
$B(E2;0_2^+\to 2_1^+)/B(E2;0_1^+\to 2^+_{2,3})$
of the inter-band transition strengths between the $K=0_1$ and $K=0_2$ bands.
If the $K=0_1$ and $K=0_2$ bands are composed of only the $K=0$ component 
and the intrinsic structures in the $(\beta,\gamma)$ plane 
are the same within the band members, this ratio should be one.
These ratios for $^{34}$Mg and $^{36}$Mg are 
close to one, indicating that the change
of the intrinsic structure between the $0^+$ and $2^+$ states is small.
In contrast, the ratios for $^{30}$Mg and $^{32}$Mg are larger than 10,
indicating a remarkable change in the shape-fluctuation properties
between the $0^+$ and $2^+$ states belonging to the $K=0_1$ and $K=0_2$ bands.

% figure 4: collective wave functions in two dimensions

\begin{figure}
\begin{tabular}{ccc}
\includegraphics[width=25mm]{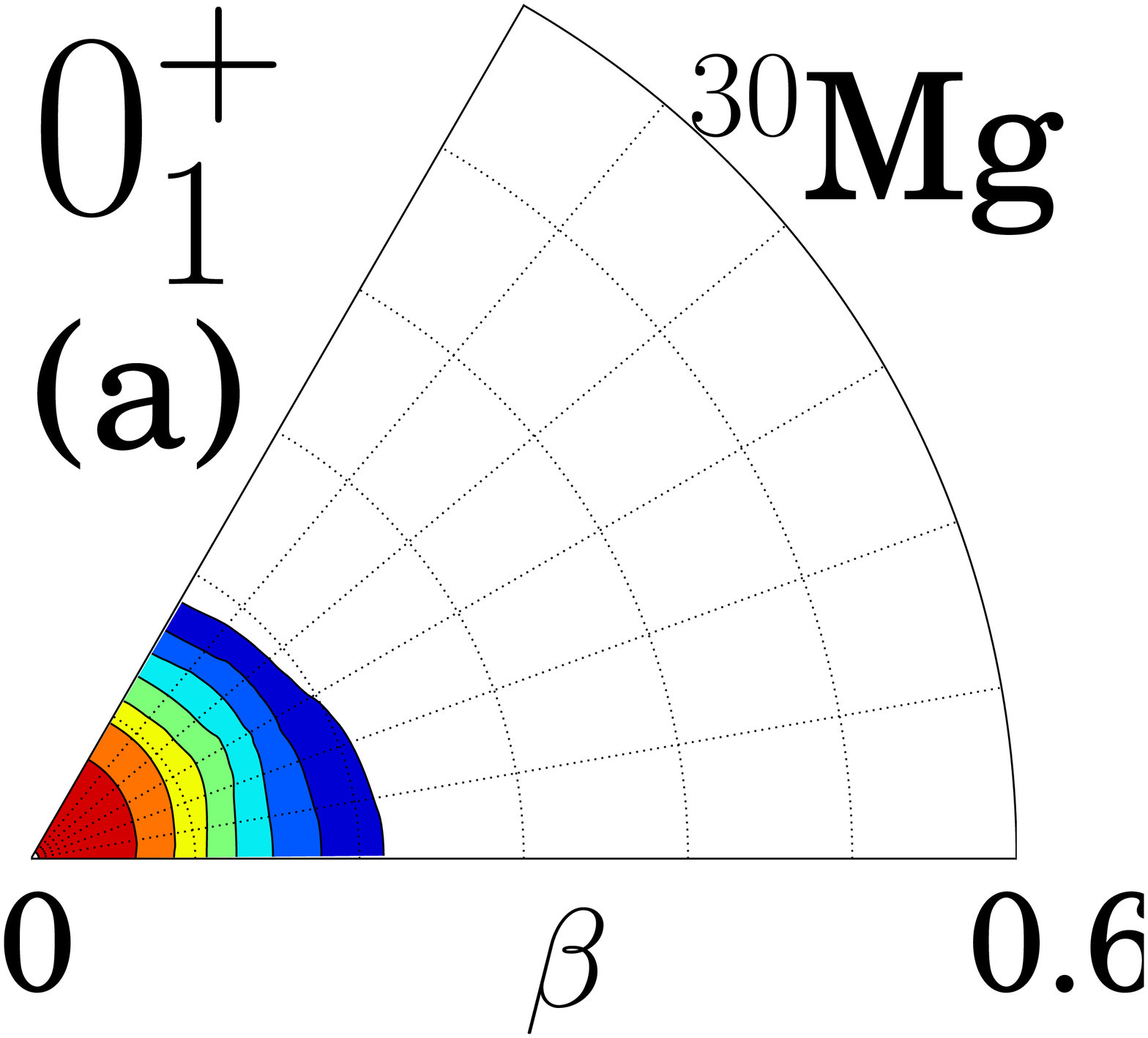} &
\includegraphics[width=25mm]{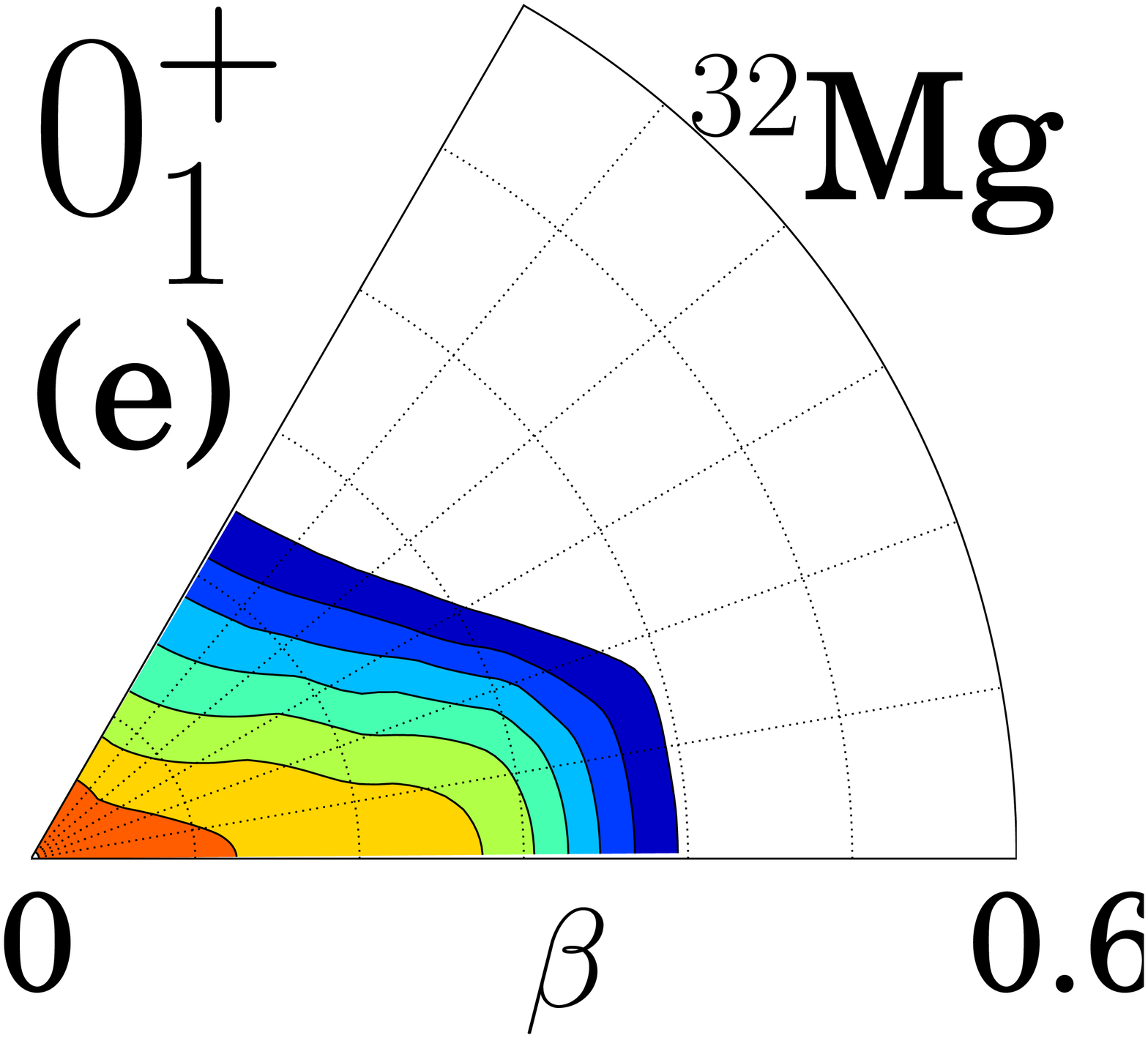} &
\includegraphics[width=25mm]{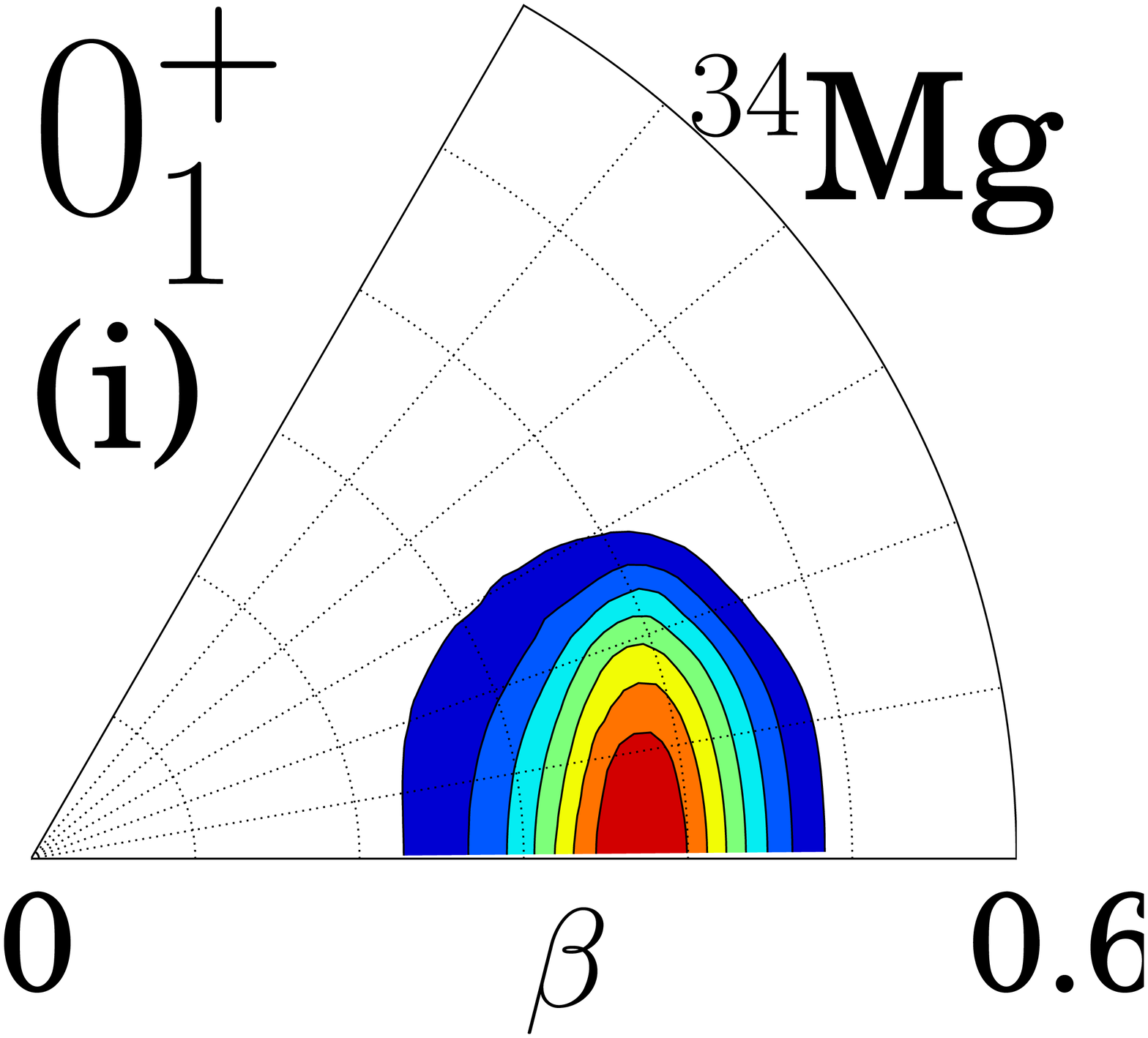} \\
\includegraphics[width=25mm]{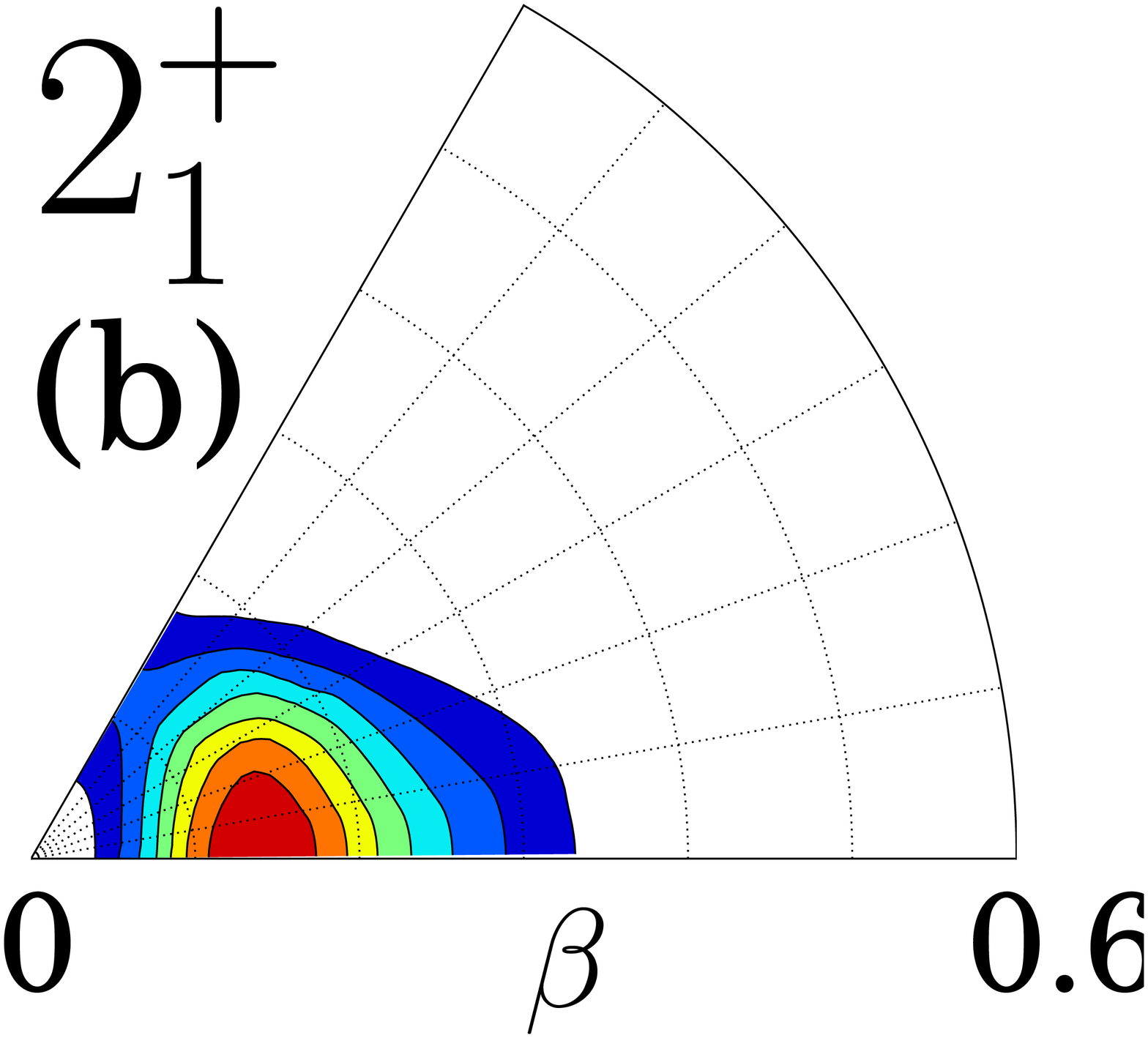} &
\includegraphics[width=25mm]{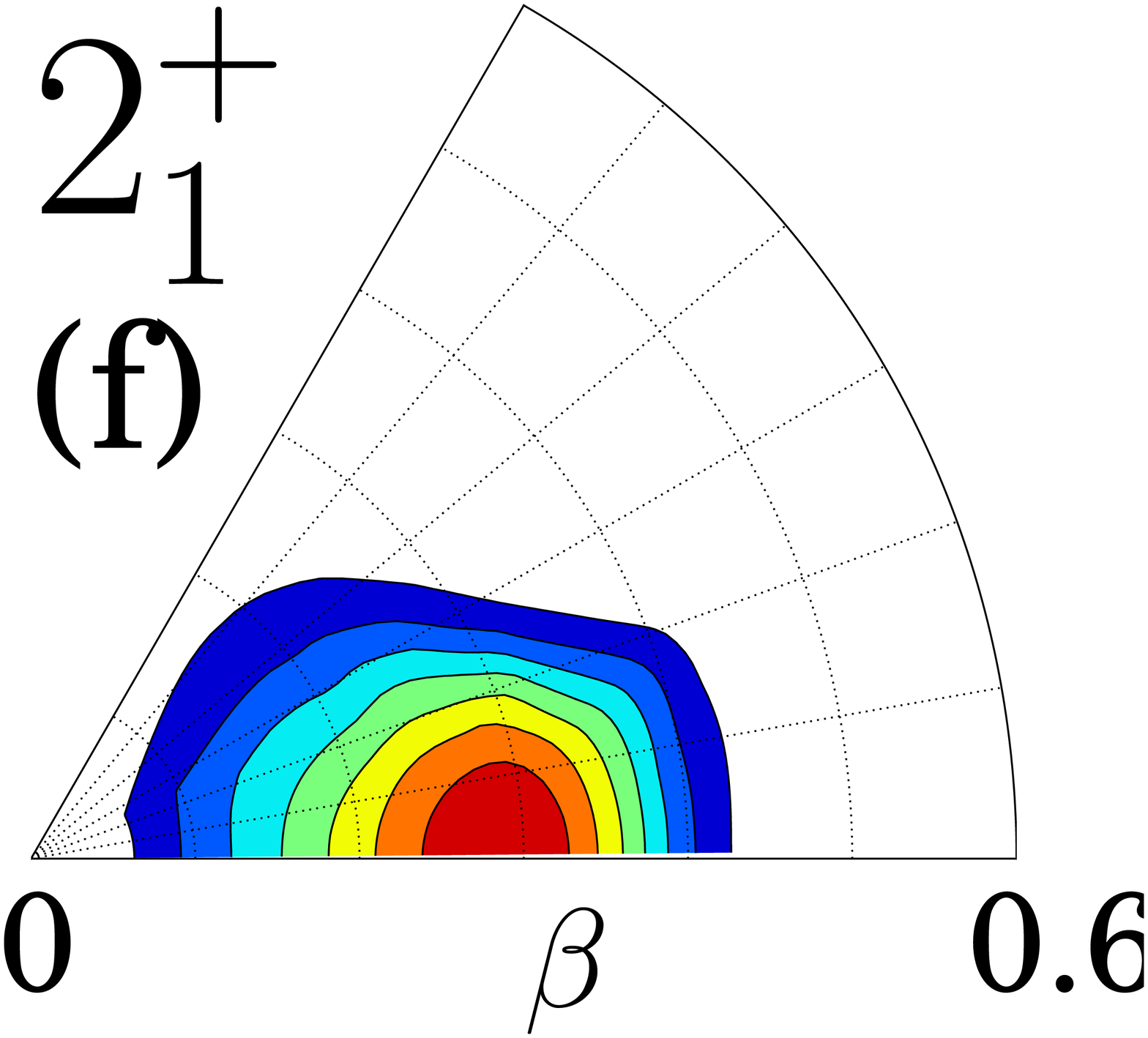} &
\includegraphics[width=25mm]{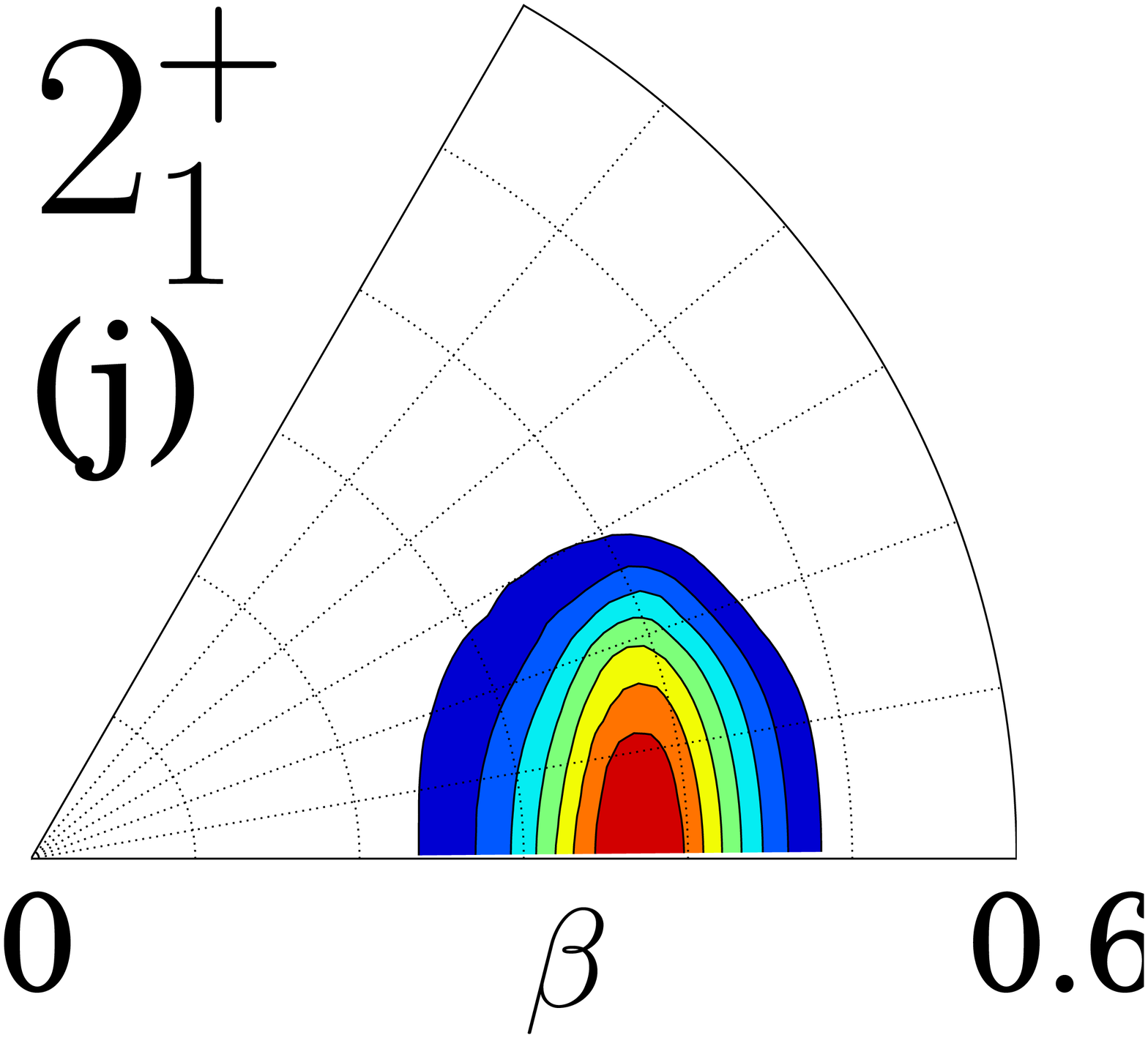} \\
\includegraphics[width=25mm]{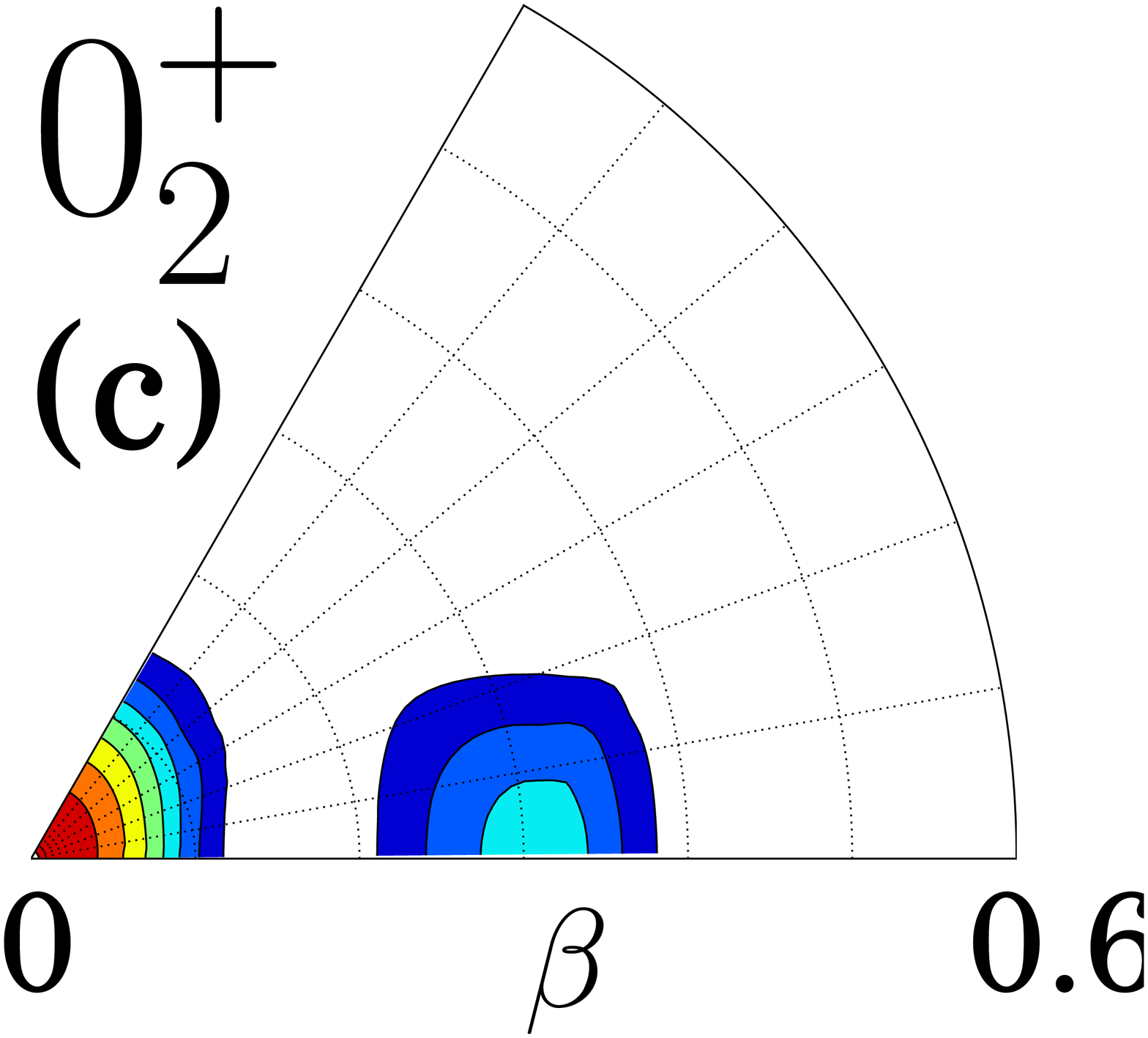} &
\includegraphics[width=25mm]{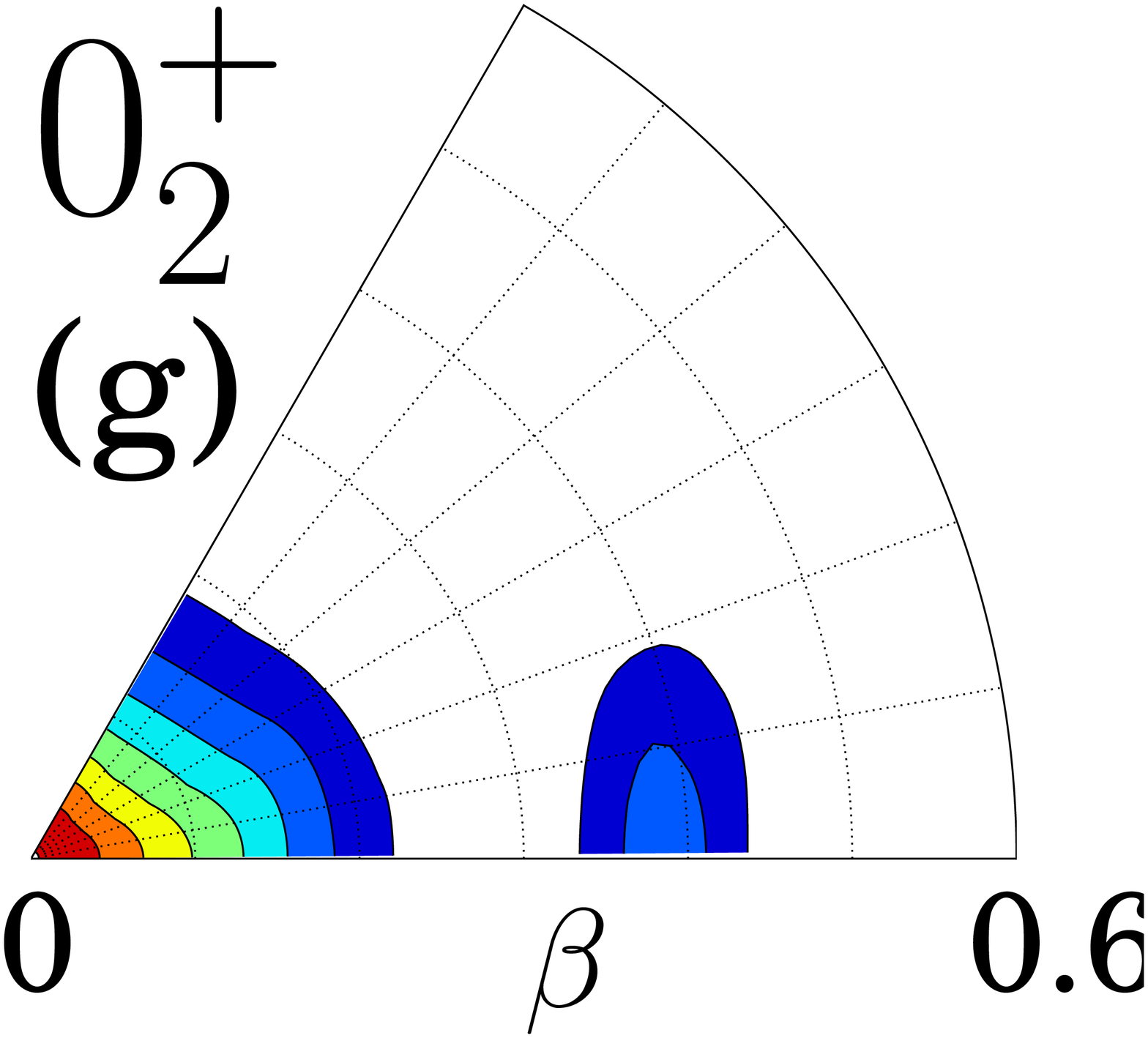} &
\includegraphics[width=25mm]{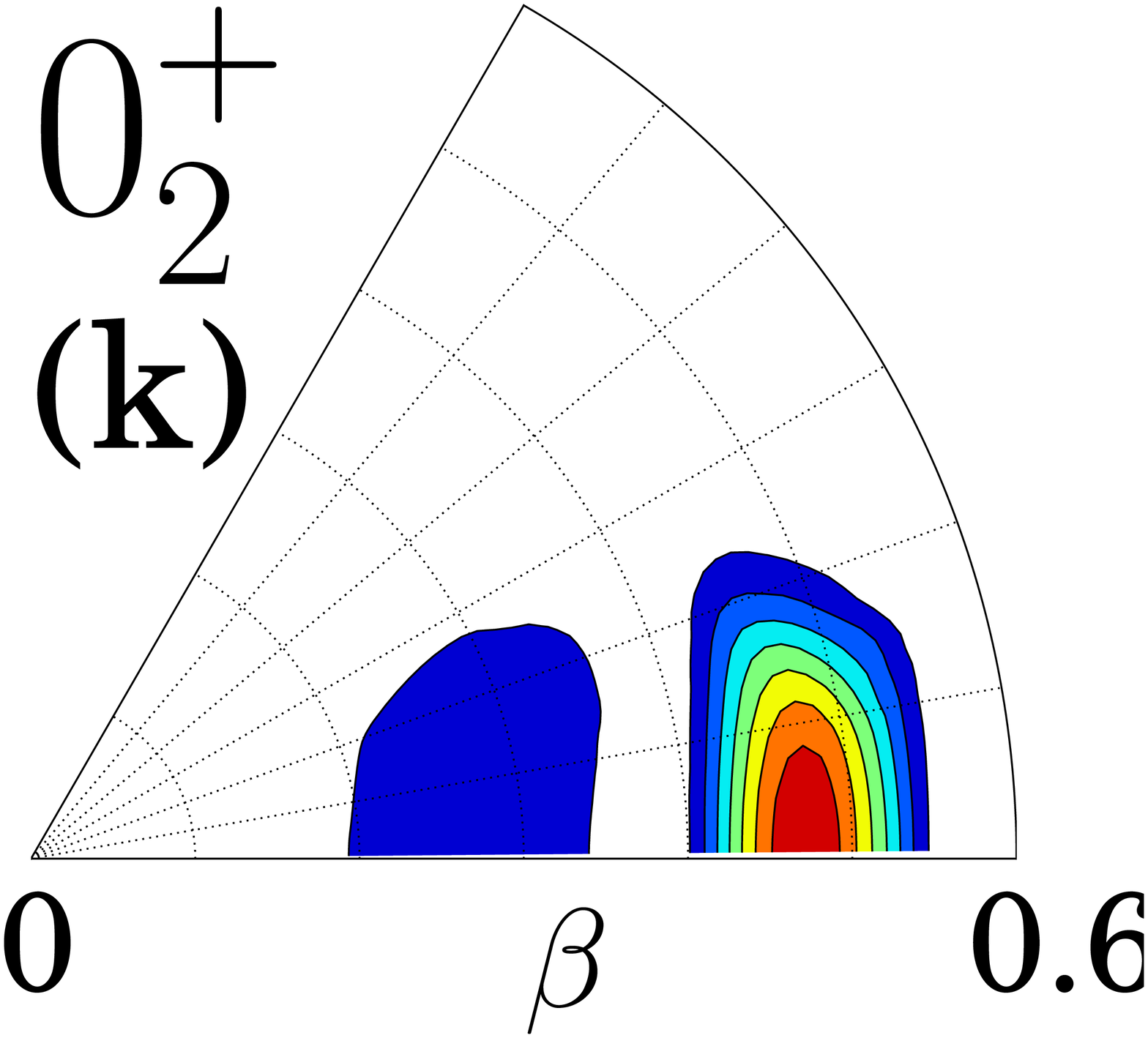} \\
\includegraphics[width=25mm]{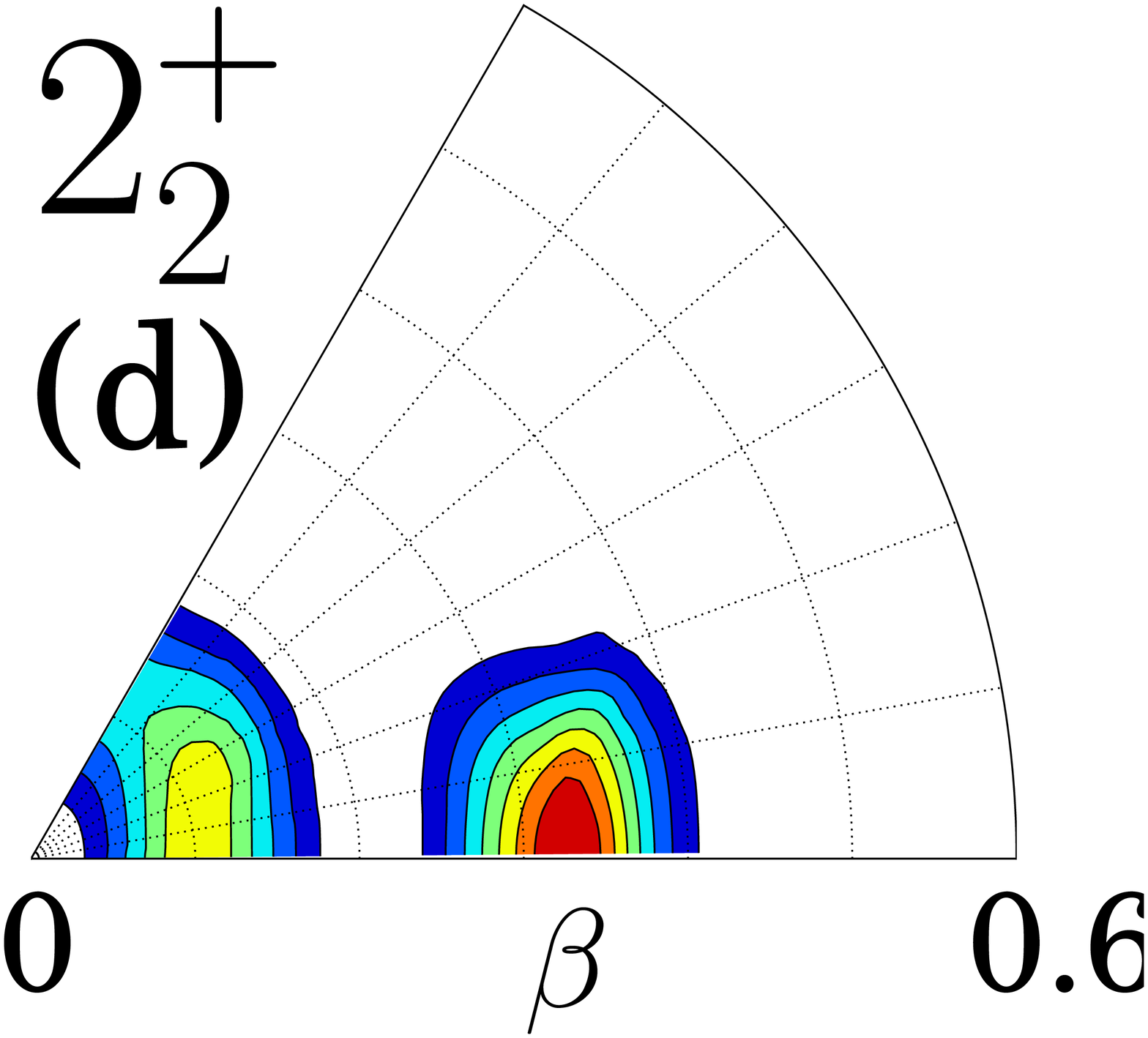} &
\includegraphics[width=25mm]{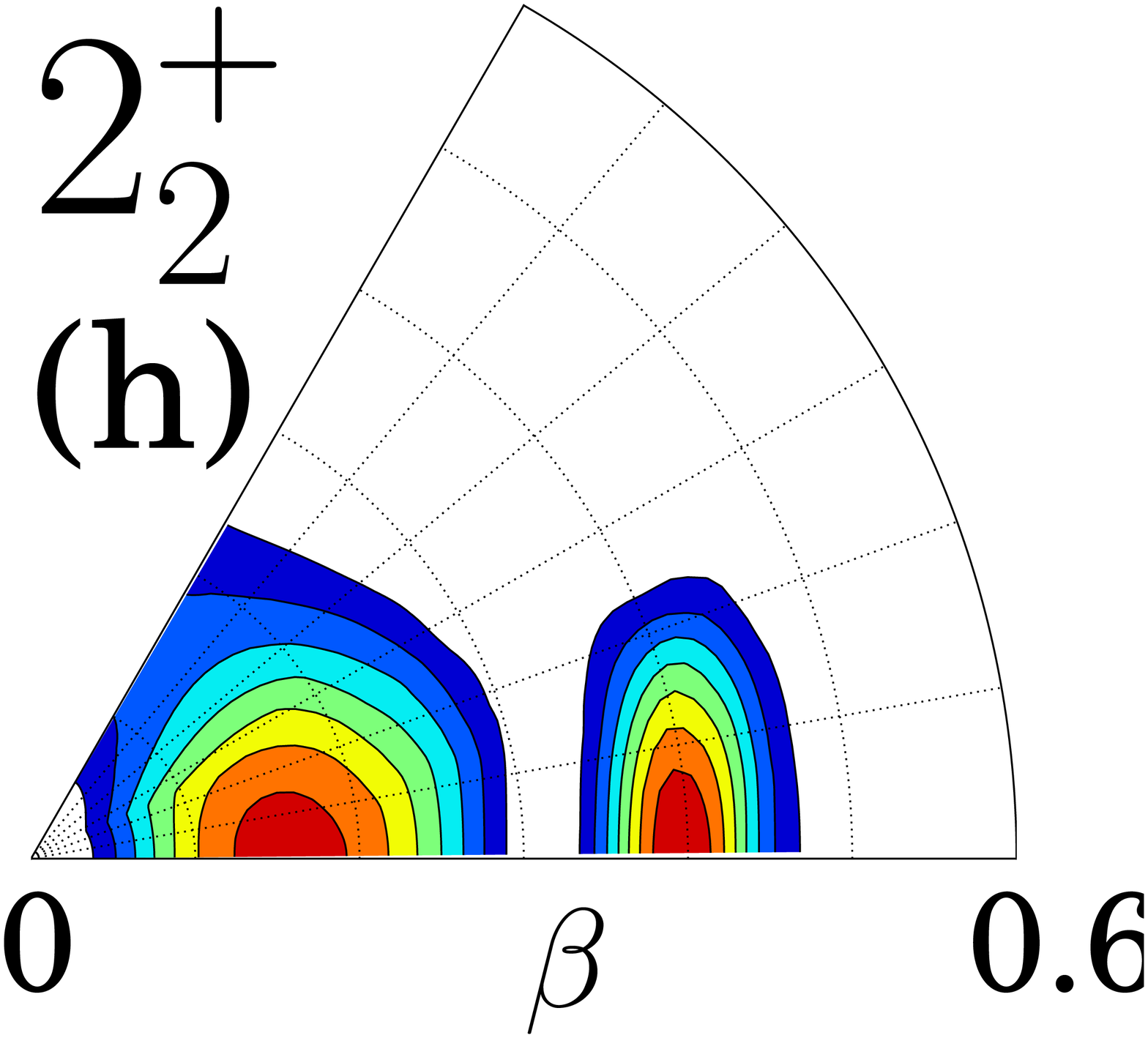} &
\includegraphics[width=25mm]{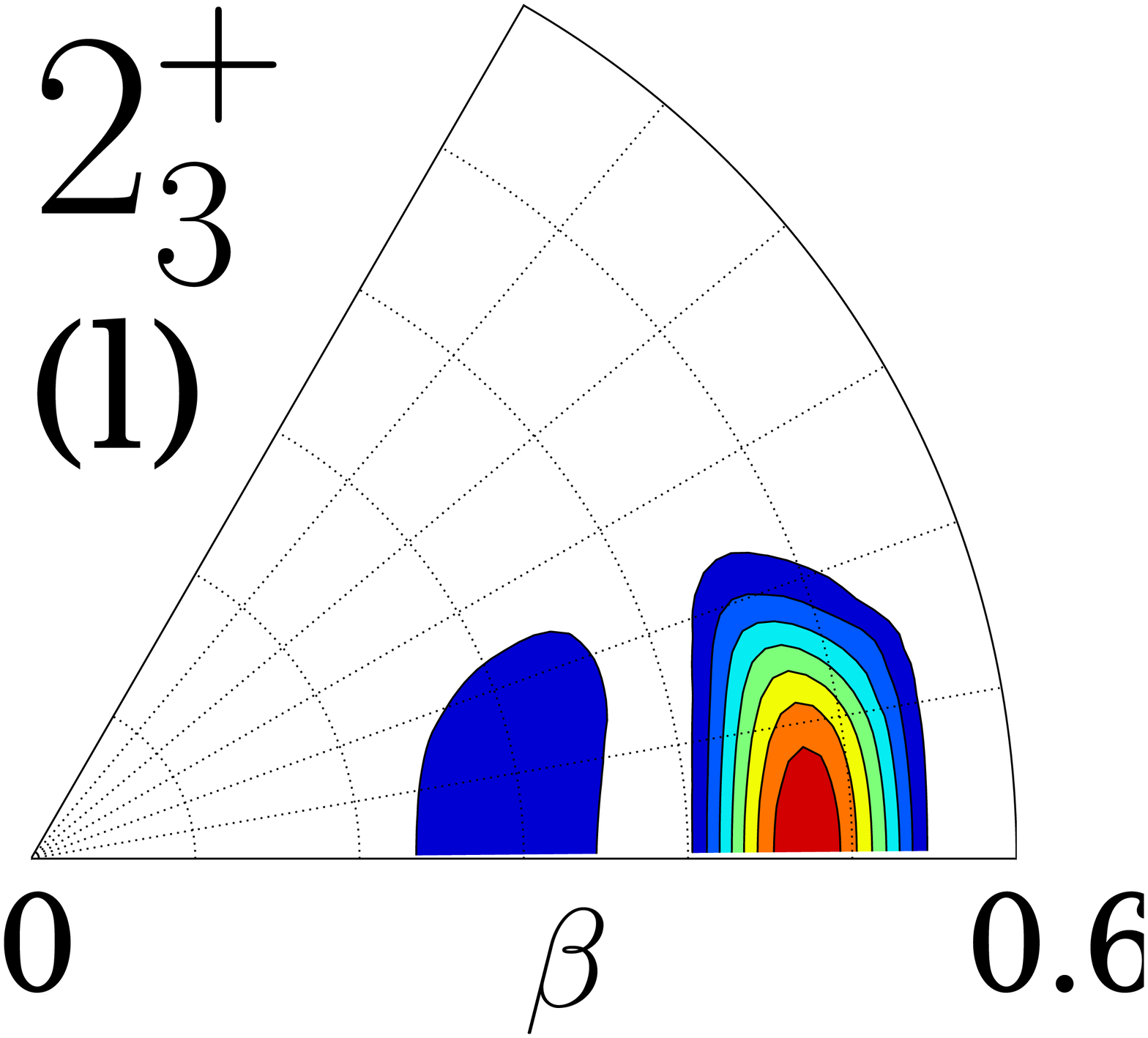} 
\end{tabular}
\caption{\label{fig:wave} (Color online)
Vibrational wave functions squared $\sum_K|\Phi_{\alpha IK}(\beta,\gamma)|^2$
of the $0_1^+, 2_1^+, 0_2^+$ and $2^+_{2,3}$ states in $^{30-34}$Mg. 
Contour lines are drawn at every eighth part of the maximum value.
}
\end{figure}

Figure~\ref{fig:wave} shows the vibrational wave functions squared $\sum_K |\Phi_{\alpha IK}(\beta,\gamma)|^2$.
Let us first examine the character change of the 
ground state from $^{30}$Mg to $^{34}$Mg. 
In $^{30}$Mg, the vibrational wave function of the ground $0_1^+$ state is 
distributed around the spherical shape. 
In $^{32}$Mg, it is remarkably extended to the prolately deformed region.  
In $^{34}$Mg, it is distributed around the prolate shape.  
From the behavior of the vibrational wave functions, 
one can conclude that shape fluctuation in the ground $0_1^+$ state is largest in $^{32}$Mg.
To understand the microscopic mechanism of this change from $^{30}$Mg to $^{34}$Mg, 
it is necessary to take into account not only the properties of the collective potential
in the $\beta$ direction but also its curvature in the $\gamma$ direction and 
the collective kinetic energy (collective masses). 
This point will be discussed in our forthcoming full-length paper. 
As suggested from the behavior of the inter-band $B(E2)$ ratio, 
the vibrational wave functions of the $2_1^+$ state are noticeably different 
from those of the $0_1^+$ state in $^{30}$Mg and $^{32}$Mg, 
while they are similar in the case of $^{34}$Mg. 
Next, let us examine the vibrational wave functions 
of the $0_2^+$ and $2_{2,3}^+$ states in $^{30-34}$Mg. 
It is immediately seen that they exhibit one node in the $\beta$ direction. 
This is their common feature. 
In $^{30}$Mg and $^{32}$Mg, one bump is seen  
in the spherical to weakly-deformed region,  
while the other bump is located in the prolately deformed region around $\beta=0.3-0.4$.
In $^{34}$Mg, the node is located near the peak of  
the vibrational wave function of the $0_1^+$ state,  
suggesting that they have $\beta$-vibrational properties. 

% figure 5: collective wave functions in one dimension

\begin{figure}[htbp]
\begin{flushleft}
\includegraphics[width=90mm]{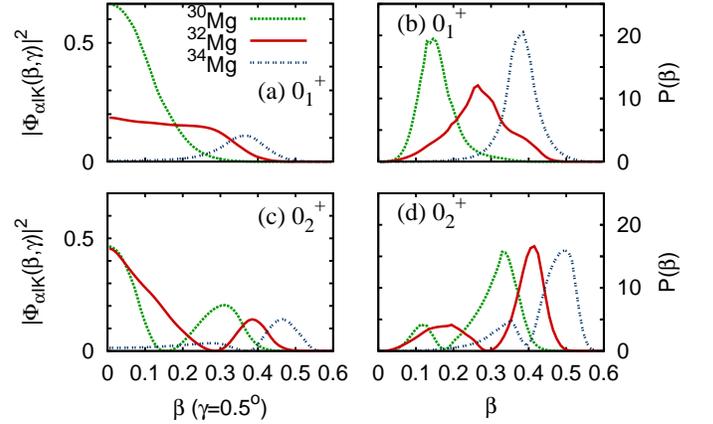}
\end{flushleft}
\caption{ \label{fig:wave1dim} (Color online)
(a) Vibrational wave functions squared,  
$|\Phi_{\alpha,I=0,K=0}(\beta,\gamma=0.5^\circ)|^2$,  
of the $0_1^+$ states in $^{30-34}$Mg. 
Their values along the $\gamma=0.5^\circ$ line are plotted as functions of $\beta$.
(b) Probability densities integrated over $\gamma$,  
$P(\beta) \equiv \int d\gamma |\Phi_{\alpha,I=0,K=0}(\beta,\gamma)|^2 |G(\beta,\gamma)|^{1/2}$, 
of the $0_1^+$ states in $^{30-34}$Mg, plotted as functions of $\beta$.
(c) Same as (a) but for the $0_2^+$ states.  
(d) Same as (b) but for the $0_2^+$ states.
}
\end{figure}

To further reveal the nature of the ground and excited $0^+$ states, 
it is important to examine not only their vibrational wave functions 
but also their probability density distributions. 
Since the 5D collective space is a curved space,  
the normalization condition for the vibrational wave functions is given by 
\begin{align}
  \int \sum_{K} |\Phi_{\alpha IK}(\beta,\gamma)|^2 |G(\beta,\gamma)|^{1/2} d\beta d\gamma =1 
\end{align} 
with the volume element 
\begin{align}
|G(\beta,\gamma)|^{1/2} d\beta d\gamma = 2 \beta^4 \sqrt{ W(\beta,\gamma)
R(\beta,\gamma)} \sin 3\gamma d\beta d\gamma,
\end{align}\begin{align}
W(\beta,\gamma) =& \{ 
D_{\beta\beta}(\beta,\gamma) D_{\gamma\gamma}(\beta,\gamma) - [D_{\beta\gamma}(\beta,\gamma)]^2
\} \beta^{-2}, \\
R(\beta,\gamma) =& D_1(\beta,\gamma) D_2(\beta,\gamma) D_3(\beta,\gamma),
\end{align}
where $D_{k=1,2,3}$ are the rotational masses defined through 
${\cal J}_k = 4\beta^2 D_k \sin^2 (\gamma - 2\pi k/3)$.
Thus, the probability density 
of taking a shape with specific values of ($\beta, \gamma$) is given by 
$\sum_K|\Phi_{\alpha IK}(\beta,\gamma)|^2 |G(\beta,\gamma)|^{1/2} $. 
Due to the $\beta^4$ factor in the volume element, 
the spherical peak of the vibrational wave function disappears  
in the probability density distribution. 
Accordingly, it will give us a picture quite different from that of the wave function.  
Needless to say, it is important to examine both aspects to understand 
the nature of individual quantum states. 

In Fig.~\ref{fig:wave1dim}, we display the probability density integrated over $\gamma$, 
$P(\beta)\equiv \int d\gamma |\Phi_{\alpha,I=0,K=0}(\beta,\gamma)|^2 |G(\beta,\gamma)|^{1/2}$, 
of finding a shape with a specific value of $\beta$, 
together with the vibrational wave functions squared 
$|\Phi_{\alpha,I=0,K=0}(\beta,\gamma)|^2$ for the ground and excited $0^+$ states ($\alpha=1$ and 2). 
Let us first look at the upper panels for the ground states. 
We note that, as expected,  the spherical peak of the vibrational wave function
for $^{30}$Mg in Fig.~\ref{fig:wave1dim}(a) corresponds 
to the peak at $\beta \simeq 0.15$ of the probability density in Fig.~\ref{fig:wave1dim}(b). 
In Fig.~\ref{fig:wave1dim}(b), 
the peak position moves toward
a larger value of $\beta$ in going from $^{30}$Mg to $^{34}$Mg.
The distribution for $^{32}$Mg is much broader 
than those for $^{30}$Mg and $^{34}$Mg. 

Next, let us look at the lower panels in Fig.~\ref{fig:wave1dim} for the excited states. 
In Fig.~\ref{fig:wave1dim}(c). 
the vibrational wave functions for $^{30}$Mg and $^{32}$Mg   
exhibit the maximum peak at the spherical shape. 
However, these peaks become small and are shifted to the region with $\beta \simeq 0.1$ and   
$\beta \simeq 0.2$ in $^{30}$Mg and $^{32}$Mg, respectively, 
in Fig.~\ref{fig:wave1dim}(d).
On the other hand, the second peaks at $\beta \simeq 0.3$ and 
$\beta \approx 0.4$ in $^{30}$Mg and $^{32}$Mg, respectively, 
seen in Fig.~\ref{fig:wave1dim}(c) become the prominent peaks 
in Fig.~\ref{fig:wave1dim}(d). 
In $^{30}$Mg, the bump at $\beta \simeq 0.1$ is much smaller 
than the major bump around $\beta \simeq 0.3$. 
In this sense, we can regard the  $0_2^+$ state of $^{30}$Mg 
as a prolately deformed state.  
In the case of $^{32}$Mg, the probability density exhibits a very broad distribution 
extending from the spherical to deformed regions up to $\beta=0.5$ 
with a prominent peak at $\beta \simeq 0.4$
and a node at $\beta \simeq 0.3$. 
The position of the node coincides with the peak of the probability density distribution 
of the the $0_1^+$ state, as expected from the orthogonality condition. 
The range of the shape fluctuation of the $0_2^+$ state in $\beta$ direction
is almost the same as that of  the $0_1^+$ state. 
Thus, the result of our calculation yields a physical picture for the $0_2^+$ state 
in $^{32}$Mg that is quite different from the `spherical excited $0^+$ state' 
interpretation based on the inversion picture of the spherical and deformed configurations.  
In $^{34}$Mg, the peak is shifted to the region with a larger value of $\beta$ 
and the tail toward the spherical shape almost disappears. 

In summary, we have investigated 
the large-amplitude collective dynamics in the low-lying states of $^{30-36}$Mg 
by solving the 5D quadrupole collective Schr\"odinger equation. 
The collective masses and potentials of the 5D collective Hamiltonian are
microscopically derived with use of the CHFB + LQRPA method.  
Good agreement with the recent experimental data is obtained
for the excited $0^+$ states as well as the ground bands.
For $^{30}$Mg, the shape coexistence picture 
that the deformed excited $0^+$  state coexists with the spherical ground state 
approximately holds. 
On the other hand, large-amplitude quadrupole-shape fluctuations 
dominate in both the ground and the excited  $0^+$  states in $^{32}$Mg, 
so that the interpretation of `deformed ground and spherical excited $0^+$ states'
based on the simple inversion picture of the spherical and deformed configurations  
does not hold.
To test these theoretical predictions,   
experimental search for the distorted rotational bands 
built on the excited $0_2^+$ states in $^{30}$Mg and $^{32}$Mg 
is strongly desired. 

One of the authors (N. H.) is supported by the Special Postdoctoral 
Research Program of RIKEN.
The numerical calculations were performed on the RIKEN Integrated
Cluster of Clusters (RICC).
This work is supported by KAKENHI (Nos.~21340073, 20105003, 23540234, and 23740223).

\bibliographystyle{apsrev}
\bibliography{../../../../Bibtex/paper.mac}% Produces the bibliography via BibTeX.

\end{document}